\documentclass[aps,groupedaddress,superscriptaddress,prd, twocolumn]{revtex4}
\usepackage{hyperref}

\usepackage{amsmath,amssymb,bm}
\usepackage{graphicx}
\usepackage{epstopdf}
\usepackage{amsfonts}
\usepackage{amssymb}
\usepackage{amsbsy}
\usepackage{amsmath}
\usepackage{array}
\usepackage{latexsym}
\usepackage{multirow}
\usepackage{sansmath}
\usepackage{subfigure}
\usepackage{lipsum}
\usepackage{float}
 \usepackage{bm}
\usepackage{color}
\usepackage{comment}
\usepackage{csquotes}		
\usepackage{physics}
 
\usepackage{accents}
 
\usepackage[plain]{algorithm}
\usepackage{algpseudocode}

\def\be{\begin{equation}}
\def\ee{\end{equation}}
\def\bea {\begin{eqnarray}}
\def\eea {\end{eqnarray}}
\def\nn {\nonumber}

\def\nn{\nonumber}

\makeatletter
\newcommand\makebig[2]{%
  \@xp\newcommand\@xp*\csname#1\endcsname{\bBigg@{#2}}%
  \@xp\newcommand\@xp*\csname#1l\endcsname{\@xp\mathopen\csname#1\endcsname}%
  \@xp\newcommand\@xp*\csname#1r\endcsname{\@xp\mathclose\csname#1\endcsname}%
}
\makeatother

\makebig{biggg} {3.0}
\makebig{Biggg} {3.5}
\makebig{bigggg}{4.0}
\makebig{Bigggg}{4.5}

\begin{document}
 
\title{Ising-like models on Euclidean black holes }

%
\author{Mustafa Saeed} \email{msaeed@unb.ca}
\affiliation{Department of Mathematics and Statistics, University of New Brunswick, Fredericton, NB, Canada E3B 5A3}

\author{Viqar Husain} \email{vhusain@unb.ca} 
\affiliation{Department of Mathematics and Statistics, University of New Brunswick, Fredericton, NB, Canada E3B 5A3}

  \begin{abstract}
\vskip 0.2cm
 We  study spin models on Euclidean black hole backgrounds. These resemble the Ising model, but are inhomogeneous with two parameters, the black hole mass and the cosmological constant. We use Monte-Carlo methods to study macroscopic properties of these systems for Schwarzschild and anti-deSitter black holes in four and five dimensions for spin-1/2 and spin-1. We find in every case that increasing the black hole mass causes the spins to undergo a second order phase transition from disorder to order and that the phase transition occurs at sub-Planckian black hole mass. 

\end{abstract}

\maketitle

\setlength{\arrayrulewidth}{0.1mm}
\setlength{\tabcolsep}{10pt}
\renewcommand{\arraystretch}{2}
\numberwithin{equation}{section}


\maketitle
 
\section{Introduction}

Spin models are useful tools in studying interacting degrees of freedom  in thermal equilibrium. This is primarily due to two reasons. Firstly due to their simplicity spin models can be solved either exactly or approximately by analytical or numerical techniques. Secondly if the simple spin model shares some global features (such as symmetries of the Hamiltonian and the number of spatial dimensions) with a real system (such as a lattice of interacting atoms) then some results of the spin model match the behaviour of the real system \cite{Cardy}.  Two well-studied spin systems are the Ising and Blume-Capel models \cite{Blume,Capel}.  

The Blume-Capel model is a spin-1 generalization of the Ising model with an additional self-interaction (or ``mass" term) with Hamiltonian  $H = -J\sum_{i,j;nn} s_is_j  + K \sum_i s_i^2$ where $nn$ denotes that the sum is over nearest neighbors. It has richer physics than the Ising model; in two dimensions, if $K>J$ there is no phase transition; if $K\le J$ then the ratio $K/J$  not only determines whether the phase transition is continuous (second order) or discontinuous (first order) but also fixes the critical temperature. The Blume-Capel model has been rigorously investigated using not only analytical techniques - such as renormalization group \cite{Burkh} - but by numerical Monte Carlo simulations also \cite{Sil}.


Monte Carlo studies of the Ising and Blume-Capel models typically use periodic boundary conditions. Due to this choice the lattice has the topology of a 2 torus. There has been a growing interest in investigating the effects of using different topologies such as the 2 sphere \cite{LH} and a random topology \cite{Ferna}. This interest is sourced from multiple areas, including  discrete models of quantum gravity, condensed matter physics and 
the statistical mechanics of complex networks such as the internet \cite{smcs}. There is also a rich literature on investigating the effects of geometric curvature on the thermodynamic properties of spins. For example it is known that the Ising model on a 2d hyperbolic plane with free boundary conditions exhibits multiple phase transitions \cite{Wu}. This model is expected to have applications to quantum information \cite{bpr,bvckt}. 

Our goal in this paper is to generalize the Ising and Blume-Capel models to Schwarzschild and AdS black hole backgrounds and study their physical content. This is of interest for several reasons: black hole backgrounds are inhomogeneous (i.e. have explicit dependence on radial location) and thus provide an interesting platform to generalize spin-models; Euclidean versions of black hole metrics are used in attempts to understand black hole thermodynamics \cite{bhthd,Gibbons}; black holes metrics are considered as thermal backgrounds for quantum field theory where the radius of Wick rotated time is proportional to black hole mass; the case with a cosmological constant provides a generalization of spin models where the connection between metric parameters and temperature is more involved \cite{Hawking1}. We review these features before constructing the models. 

Our approach is to start with a scalar field action on a general form of a Euclidean black hole background, discretize the action on a suitably defined lattice, and then restrict the scalar field to take on spin values. This process naturally introduces a point interaction from the scalar field mass term, and nearest neighbour interactions from a finite difference of the kinetic term.

In Section II we review the construction of the Euclidean black holes  \cite{Gibbons, Hawking1, Wald:1984rg} 
; in Section III we define the spin-models;  in Section IV we describe and present the Monte-Carlo calculations for the spin-1/2 and spin-1 models, followed by a  summary of the work and possible extensions.  

\section{Euclidean black holes}
 
\label{DESM}

As a prelude to defining spins models on Euclidean black hole backgrounds, in this section we summarize the main aspects of the metrics we will be using. 

The Schwarzschild black hole metric in its original form is  
\be
ds^2 = - \left(1-\frac{2M}{r}\right) dt^2 + \left(1-\frac{2M}{r}\right)^{-1} dr^2 + r^2 d\Omega^2,
\ee
in the coordinates $(t,r,\theta,\phi)$; it has a coordinate singularity at the horizon $r=2M$ and a curvature singularity at $r=0$. In the Kruskal-Szekeres (KS) coordinates  $(T,X, \theta,\phi)$ defined as
\bea
-T^2 + X^2 &=& \left(\frac{r}{2M}-1 \right)\exp(-\frac{r}{2M})\\
\frac{T+X}{T-X} &=& \exp(-\frac{t}{2M})
\eea
the geometry is 
\be
ds^2 = \frac{32M^3}{r}\exp\left(-\frac{r}{2M}\right)\left(-dT^2 + dX^2\right) + r^2 d\Omega^2.
\ee
In these coordinates the event horizon $r=2M$ is given by $T=\pm X$, and the singularity at $r=0$ is $T^2-X^2=1$. The KS coordinates provide a global extension of the original Schwarzschild metric, and are best suited to understanding its Euclidean version. 

The Euclidean KS metric is defined by the Wick rotation $\tau = iT$
\be
ds^2 = \frac{32M^3}{r}\exp(-\frac{r}{2M})\left(d\tau^2 + dX^2\right) + r^2 d\Omega^2,
\ee
where now $\displaystyle \tau^2 + X^2 =\left(r/2M-1\right)\exp(-r/2M)$. This rotated metric is singularity free since $r=0$ now corresponds to $\tau^2 +X^2 =-1$ which has no solution;  the horizon $r=2M$ corresponds to $\tau^2 + X^2 =0$,  which is the origin of $(\tau,X)$ plane. Thus, the Euclidean Schwarzschild geometry is singularity free and the interior of the Lorentzian Schwarzschild metric ($r< 2M$) is absent. 

Having obtained the Euclidean Schwarzschild geometry in the KS coordinates, we now see that the same result is obtained  in the original coordinates by simply restricting the radial coordinate to $r\in [2M,\infty)$, with $\tau=it$. 

The near-horizon metric  is obtained by setting $r= 2M + \mu$; then for $\mu \ll 1 $ the $\tau-\mu$ part of the metric is
\be
ds^2 = \frac{2M}{\mu}d\mu^2 + \frac{\mu}{2M} d\tau^2;
\ee
defining $d\rho^2 = (2M/\mu) d\mu^2$ gives 
\be
ds^2 = d\rho^2 + \left(\frac{\rho}{4M}\right)^2d\tau^2. 
\ee
The final observation is that absence of a conical singularity at $\rho=0$ requires $\tau/ 4M \in [0,2\pi)$.

This procedure is readily generalized  \cite{Wald:1984rg} for obtaining a regular Euclidean black hole geometry starting from the $d-$dimensional metric 
\begin{equation}
\label{eq:lmdp1}
ds^{2}=-F(r)dt^{2}+F(r)\,^{-1}dr^{2}+r^{2}d\Omega_{d-2}\,^{2},
\end{equation}
where $F(r)$ is a one to one function such that there is an $r_{0}\in [0,\infty)$ such that $F(r_0)=0$ (the horizon), $F'(r_0)>0$ and $\lim_{r\rightarrow 0} F(r)=\infty$ (the singularity). Then, 
the Wick rotation $\tau=it$, the restriction $r\in [r_0,\infty)$ and the coordinate transformation 
\be
\rho(r)=\beta_{0}\sqrt{F(r)}, \qquad \beta_{0}=\frac{2}{F'(r_0)} 
\label{eq:rrhor}
\ee
gives the metric 
\bea 
\label{eq:rrhor1}
ds^{2}&=&\left(\frac{\rho}{\beta_0}\right)^2 d\tau^2 +\left(\frac{2}{\beta_0F'(r)}\right)^2 d\rho^2 \nn\\
&&\quad +\ r^2(\rho)\ d\Omega_{d-2}\,^2,
\eea
where $\tau\in (-\infty,\infty)$ and $\rho\in [0,\lim_{r\to\infty} \,\rho(r)]$. The near-horizon $r\approx r_0$ metric is 
 \be
ds^{2}=d\rho^2 + \left(\frac{\rho}{\beta_0}\right)^2 d\tau^2  + r^2(\rho)d\Omega_{d-2}\,^2.
\ee
Thus the $\tau-\rho$ part of the metric is flat space with no conical singularity provided $\tau\beta_{0}\,^{-1}$ is an angular coordinate with period $2\pi$; this corresponds to $\tau$ having periodicity $\beta$
\begin{align}
\label{eq:etp}
    \tau\in[0,\beta), \quad \beta=2\pi\beta_{0}.
\end{align}
The inverse of the Euclidean time periodicity is the natural temperature of the spacetime 
\be
T=\left(2\pi\beta_{0}\right)^{-1}.
\ee

In summary, we will use the metric (\ref{eq:rrhor1}) with the periodically identified $\tau$ as the background for defining spin models, with the function
\be
F(r) = 1- \frac{2M}{r^{d-3}} + \left(\frac{r}{L}\right)^2;
\ee
this is the AdS black hole in $d$ spacetime dimensions where the scale $L$ is related to $d$ and the cosmological constant by $\Lambda= - (d-1)(d-2)/(2L^2)$. The cases we consider are 4d Schwarzschild ($L\rightarrow\infty$) and Schwarzschild AdS in 4d and 5d. 

\begin{figure}[!hbp]
\begin{center}
\includegraphics[width=0.8\columnwidth]{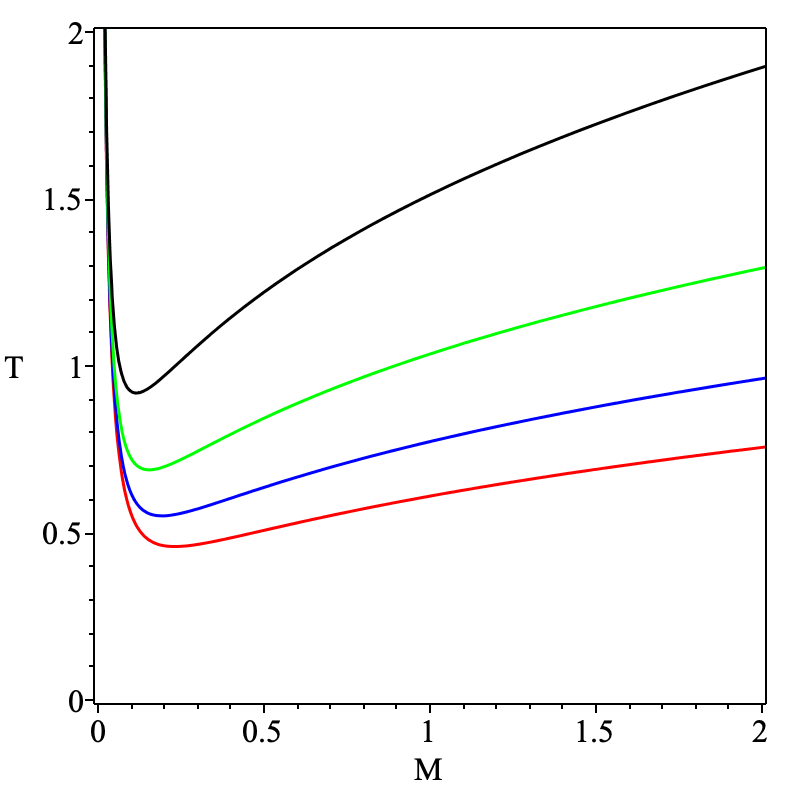}
\caption{\label{fig:AdS_TM} Temperature-Mass  relation  for Euclidean AdS-Schwarzschild  for $L=0.3,0.4,0.5,0.6$ (top to bottom).}
\end{center}
\end{figure} 

The temperature-mass relation for the 4d Schwarzschild case is $T= (4\pi r_0)^{-1}= (8\pi M)^{-1}$. However for the 4d anti-deSitter case it is 
\be
\label{MTL}
T=\frac{3r_0\,^2+L^2}{4\pi r_0L^2}; \quad M=\frac{r_0}{2}\left[1+\left(\frac{r_0}{L}\right)^2 \right].
\ee
$T(M)$ has a minimum at $\displaystyle T_{\text min}= \sqrt{3}/(2\pi L)$ at mass $\displaystyle M_*= 2L/(3\sqrt{3})$ (Fig. \ref{fig:AdS_TM}) and horizon radius $r_{0*} =L/\sqrt{3}$;  $T> T_{\text min}$  corresponds to either a small or a large mass black hole; a similar result holds the 5d Schwarzschild AdS case also. This discussion will be relevant below where we discuss phase transitions in spin models as functions of $T$ and $M$.

\section{Spin models on  black holes}

Spin models on  Euclidean black hole  backgrounds may be defined by starting with the  Euclidean action of a scalar field  on the metrics reviewed in the last section, discretizing the action, and then restricting the scalar field  variable to take discrete values, $\pm 1$ for Ising,  and $0,\pm 1$ for Blume-Capel models. The $d$-dimensional Euclidean action for a massive scalar field on a background metric $g_{ab}$   is  
\begin{equation}
\label{eq:sfa}
I_{E}\left[\Phi\right]= \frac{1}{2}\int d^{d}x\ \sqrt{g}\left(  g^{ab}\partial_{a}\Phi\partial_{b}\Phi+m^2 \Phi^2\right).
\end{equation} 
The Euclidean black hole metrics (\ref{eq:rrhor1}) have the form  
\be 
\label{eq:3Eucmetric}
ds^{2}=u(\rho)d\tau^{2}+v(\rho)d\rho^{2}+w(\rho)d\Omega_{d-2}\,^{2}.
\ee
We consider only the $\tau-\rho$ plane  since the sphere is not relevant for the  black hole properties of interest. Therefore the scalar field we consider is $\Phi\equiv\Phi(\tau,\rho)$ with the Euclidean time periodicity $\Phi(0,\rho) = \Phi(\beta,\rho)$.
With these assumptions the Euclidean action reduces to 
 \bea
\label{eq:conttodisc}
I_{E}\left[\Phi\right] 
&=& \frac{A_{d-2}}{2} \int^\beta_0\int_0^{\rho_0} d\tau d\rho \sqrt{w^{d-2}} \nn\\
 &&\cdot\left(\sqrt{\frac{v}{u}} {\dot{\Phi}}^{2}+\sqrt{\frac{u}{v}}{\Phi'}^{2}+m\sqrt{uv} 
  \Phi^{2}\right)
\eea
where $A_{d-2}$ is the area of $S^{d-2}$ and dots and primes denote partial derivatives with respect to $\tau$ and $\rho$; the $\rho$ integration is restricted to a finite value $\rho_0$ to define a discrete model on a finite lattice. The object of interest is the partition function 
\begin{equation}
Z=\int{D\Phi \exp\left(-I_{E}\left[\Phi\right]\right)}.
\end{equation}
 
\subsection{Discretization}

Consider the $\tau-\rho$ plane as an $N_{\beta}\times N_{\rho}$ lattice with spacing 
\be
\label{eq:dsin}
\epsilon = \frac{\beta}{N_{\beta}} =\frac{\rho_{0}}{N_{\rho}}.
\ee
Through $\beta$ the lattice spacing depends on the metric parameters $M$ and $\Lambda$, and a simple discretization is given by   
\bea
&&\tau\rightarrow \tau_{m}=m\epsilon,\qquad \rho \rightarrow \rho_{n} = n\epsilon;\nn\\
&&f(\tau,\rho)\rightarrow f_{m,n},\nn\\
&&\dot{f} \rightarrow  \frac{f_{m+1,n}-f_{m,n}}{\epsilon},\nn\\
&&f' \rightarrow  \frac{f_{m,n+1}-f_{m,n}}{\epsilon}.
\eea
 Lastly, the scalar field values are restricted to the set
\be
\label{eq:dsfi}
\Phi=\{-s,\, -s+1, \, ... \, \,s-1, \,s\} \times s^{-1}, \quad s\in \mathbb{N},
\ee
so that the values of $\Phi$ are rational numbers in the interval $[-1,1]$. With this discretization the Euclidean action becomes the sum   
\begin{widetext}
\bea
\label{eq:imcb}
I_{E}\left[\Phi\right]&=&A_{d-2}\sum_{m,n=1}^{N_{\beta},N_{\rho}} 
\left\{-\Phi_{m,n}\left[\sqrt{\frac{v_{n}w_{n}\,^{d-2}}{u_{n}}}\,\Phi_{m+1,n}+\sqrt{\frac{u_{n}w_{n}\,^{d-2}}{v_{n}}}\,\Phi_{m,n+1}\right] \right. \nn\\
&&\left. +\,\Phi^2_{m,n}\left[\sqrt{\frac{v_{n}w_{n}\,^{d-2}}{u_{n}}}+\frac{1}{2}\left(\sqrt{\frac{u_{n}w_{n}\,^{d-2}}{v_{n}}}+\sqrt{\frac{u_{n-1}w_{n-1}\,^{d-2}}{v_{n-1}}}\right)+\frac{m\epsilon^{2}}{2}\sqrt{u_{n}v_{n}w_{n}\,^{d-2}} \right] \right\}.
\eea
\end{widetext}
This is the desired ``spin" model. It is apparent that the coupling ``constants" vary across the lattice and depend on the metric functions from point to point; the first line contains metric function-weighed nearest neighbour interactions in the $\tau$ and $\rho$ directions; the second line represents spin self-interactions which are either weighed by the metric functions or by the determinant of the metric, scalar field mass and $\epsilon^2$ (a factor that cancels out for the remaining terms); the choice $s=1/2$ or $s=1$ gives an Ising or Blume Capel like model on Euclidean black hole background respectively; the self interaction terms only affect statistical averages in the latter case. 

For the 4d Euclidean AdS-Schwarzschild geometry the metric functions are 
\begin{align}
u(\rho)&=\left[\left(\frac{3r_{0}{}^{2}+L^{2}}{2r_{0}L^{2}}\right)\rho\right]^{2}\\
v(\rho)&=\left[\left(\frac{3r_{0}{}^{2}+L^{2}}{2r_{0}L^{2}}\right)\left(\frac{L^{2}\,r^2(\rho)}{r^3(\rho)+ML^2}\right)\right]^{2}\\
w(\rho)&=r^2(\rho),
\end{align}
where $r(\rho)$ is given by (\ref{eq:rrhor}) with $F(r)=(1-r/2M + r^2/L^2)$; it is readily verified that for this $F$ there is a single real root for $r(\rho)$, a fact that leads to unambiguous  values of $w(\rho)$.  On the lattice these functions are discretized with the replacements $\rho \rightarrow \rho_n = n\epsilon$ and $r_n(\rho_n)$. This results in the specific spin model 
\begin{widetext}
\begin{equation}
\begin{split}
I_{E}\left[\Phi\right]=&\ 4\pi\sum_{m,n=1}^{N_{\beta},N_{\rho}}\,\Bigg\{-\Phi_{m,n}\left[\left(\frac{r_{n}{}^{4}}{\rho_{n}}\right)\left(\frac{L^{2}}{ML^{2}+r_{n}{}^{3}}\right)\Phi_{m+1,n}+\left(\rho_{n}\right)\left(\frac{ML^{2}+r_{n}{}^{3}}{L^{2}}\right)\Phi_{m,n+1}\right]\\
&+\left[\left(\frac{r_{n}{}^{4}}{\rho_{n}}\right)\left(\frac{L^{2}}{ML^{2}+r_{n}{}^{3}}\right)+\left(\frac{ML^{2}+r_{n}{}^{3}}{2L^{2}}\right)\left(\rho_{n}+\rho_{n-1}\right)\right]\Phi_{m,n}\,^{2}\\
&+\left(\frac{\rho_{n} r_{n}{}^{4}L^{2}}{ML^{2}+r_{n}{}^{3}}\right)\left(\frac{3r_{0}{}^{2}+L^{2}}{2r_{0}L^{2}}\right)^{2}\,\epsilon^{2}\,\left(\frac{m}{2}\right)\Phi_{m,n}\,^{2}\Bigg\}.
\end{split}
\label{eq:imsa4}
\end{equation}
The limit $L\rightarrow \infty$ gives the model on 4d Euclidean Schwarzschild background:
\begin{equation}
\label{eq:imsb}
\begin{split}
I_{E}\left[\Phi\right]=&4\pi\sum_{m,n=1}^{N_{\beta},N_{\rho}}\,\Biggg\{-\Phi_{m,n}\left[\frac{16M^{3}}{\rho_{n}\left(1-\left(\frac{\rho_{n}}{4M}\right)^{2}\right)^{4}}\,\,\Phi_{m+1,n}+\left(\rho_{n} M\right)\Phi_{m,n+1}\right]\\
&+\left[\frac{16M^{3}}{\rho_{n}\left(1-\left(\frac{\rho_{n}}{4M}\right)^{2}\right)^{4}}+\frac{M}{2}\left(\rho_{n}+\rho_{n-1}\right)\right]\Phi_{m,n}\,^{2}+\left[\frac{M\rho_{n}}{\left(1-\left(\frac{\rho_{n}}{4M}\right)^{2}\right)^{4}}\right]\,\epsilon^{2}\,\left(\frac{m}{2}\right)\Phi_{m,n}\,^{2}\Biggg\}.
\end{split}
\end{equation}
\end{widetext}
A similar procedure gives the model on the 5d AdS-Schwarzschild metric, and indeed any metric of the form (\ref{eq:3Eucmetric}).

This completes the prescription for defining spin models on Euclidean black hole  spaces. The process we have followed is similar to discretizing any continuum theory on a lattice where certain discretization choices are made; in our case this is the representation for time and space derivatives. The procedure we have followed is analagous to discretized thermal quantum field theory (TQFT) on Euclideanized Minkowski spacetime with periodic identification of Wick rotated time. But there are important differences: in the black hole case the metric comes with the mass and cosmological constant parameters, and the periodicity of Euclidean time (which is guided by the requirement to remove the conical singularity) depends on these parameters.

\section{Simulation details and results}

As we have noted, in accordance with the relation 
\eqref{eq:etp} between the $\tau$ dimension size and black hole mass, the lattice size in the $\tau$ direction varies with $M$. Thus the thermodynamics properties of spins are best studied as a function of $M$ with fixed lattice spacing $\epsilon$ but varying $\tau$-dimension lattice size.  We perform Monte Carlo simulations with $\epsilon=0.01$, and take 
\be
N_\beta = \left\lfloor \frac{\beta}{\epsilon} \right\rfloor, \quad 
 N_\rho = \left\lfloor \frac{\rho_0}{\epsilon} \right\rfloor
\ee
where $\beta$ is given by  (\ref{eq:etp}) and we take $\rho_0 = \rho(5r_0)$ as the radial extent of the lattice. For convenience we use periodic boundary conditions in the $\rho$ direction and $m=0$ for all simulations in addition to $\Lambda =-4$ for the AdS simulations. 

After choosing a particular Euclidean black hole metric (Schwarzschild, 4d Schwarzschild-AdS or 5d Schwarzschild-AdS), we select a set of $M$ values; each value of $M$ fixes the lattice size. We use an aligned lattice (a cold start) once an $M$ value is selected. Subsequent configurations are generated according to the probability distribution $\displaystyle 
P(x)=\exp[-I_{E}(x)]$ using a Monte Carlo Markov Chain (MCMC). We use the standard  procedure where a single Monte Carlo (MC) step entails flipping a randomly selected spin, computing  the resulting change in the action $\Delta I_E$, and keeping the new configuration provided  $\displaystyle \exp\left(-\Delta I_{E}\right) \geq \text{rand}[0,1]$; a sweep is defined as $N_{\beta}\times N_{\rho}$ MC steps. 

\begin{figure}[!hbp]
\begin{center}
\includegraphics[width=0.8\columnwidth]{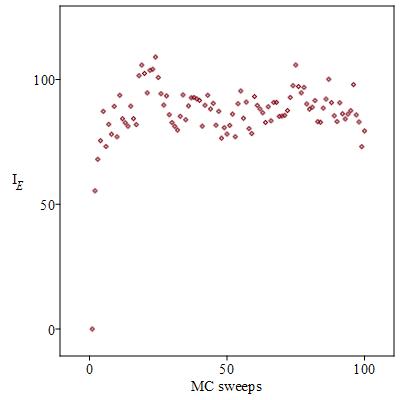}
\caption{\label{fig:Therm}Spins on Euclidean Schwarzschild background with $M=0.025$ are thermalized at around the $50^{th}$ MC sweep.}
\end{center}
\end{figure}

Thermalization requires a number of sweeps from the starting cold configurations. This is determined by computing $I_E$ as a function of the number of sweeps until a steady state is reached. Fig. \ref{fig:Therm} shows that the required number of sweeps to thermalization is about 50 for the Euclidean Schwarzschild case with $M=0.025$
. 

After thermalization we compute the average values of several thermodynamic quantities ${\cal O}$ with an additional $N_M=2000$ sweeps for each value of mass $M$ on its corresponding lattice using the formula 
\be
\langle {\cal O}\rangle = \frac{1}{N_M}\ \sum_{k=1}^{N_M} {\cal O}_i.
\ee
Here $i$ is the number of the sweep and $\mathcal{O}_{i}$ is the corresponding measurement of the thermodynamic quantity. The specific quantities ${\cal O}$ we calculate include the alignment, energy, susceptibility and  specific heat:
\bea
A&=&\left(\frac{1}{N_{\beta}N_{\rho}}\right)\,\Bigg|\,\sum_{m,n=1}^{N_{\beta},N_{\rho}}\Phi_{m,n}\,\Bigg|,\\
E&=&\left(\frac{1}{N_{\beta}N_{\rho}}\right)\,I_{E},\\
\chi &=& \frac{1}{T}\ \left(\langle A^2\rangle - \langle A \rangle^2 \right),\\
C&=&\frac{1}{T^2}\ \left(\langle E^2\rangle - \langle E \rangle^2 \right).
\eea

Another observable we calculate is entropy $S(T)$. A computation of $S(T)$ without using the partition function $Z$ (which is not available) is to utilize a discrete version of the formula
\be
S(T) = \int_0^T \frac{C(T')}{T'} dT'.
\ee
A discrete estimate is 
\be
S(T)\approx\sum_{i=1}^{N_{T}} \left[\frac{C(T_{i})}{T_{i}}\right]\left(T_{i}-T_{i-1}\right),
\ee
where $N_{T}$ is the total number of temperatures simulated till $T$ in MC runs, and $T_{i}$ represents the $i^{th}$ temperature in that list (where $T_1$ is the minimum temperature possible due to computational or theoretical constraints) and $C(T_{i})$ is the  heat capacity numerically evaluated at $T_{i}$. 

For computing $A$, $\chi$, $C$ and $S$, let us recall that a lattice is determined by $M$, and for the AdS black holes a value of $M$ uniquely determines $T$, but not vice versa (Fig. 1). Therefore these quantities as functions of $M$ are more naturally visualized for the AdS cases. However it is also useful to see these quantities as functions of $T$; for a given $T$ there are two values of $M$, the small and large black holes; these correspond to two different lattices, one for each mass. Therefore we expect two sets of results for each of $A$, $\chi$, $C$ and $S$ as functions of $T$,  one for the small black hole and the other for the large black hole. 

\subsection{Numerical results}

We computed the alignment, susceptibility, specific heat, and entropy    for spin-1/2 and spin-1 models for Schwarzschild, and 4d and 5d Schwarzschild AdS. In each case we find  evidence of a phase transition at a critical value of mass $M_c$; in each case it turns out that $M_c$ is sub-Planckian with value indicated in the figures. We first plot these quantities as functions of $M/M_c$, and again as functions of $T/T_c$, where $T_c$ for the AdS cases is uniquely determined (as for $\Lambda=0$) by $M_c$. 

\begin{figure*}
\begin{center}
\includegraphics[width=\textwidth]{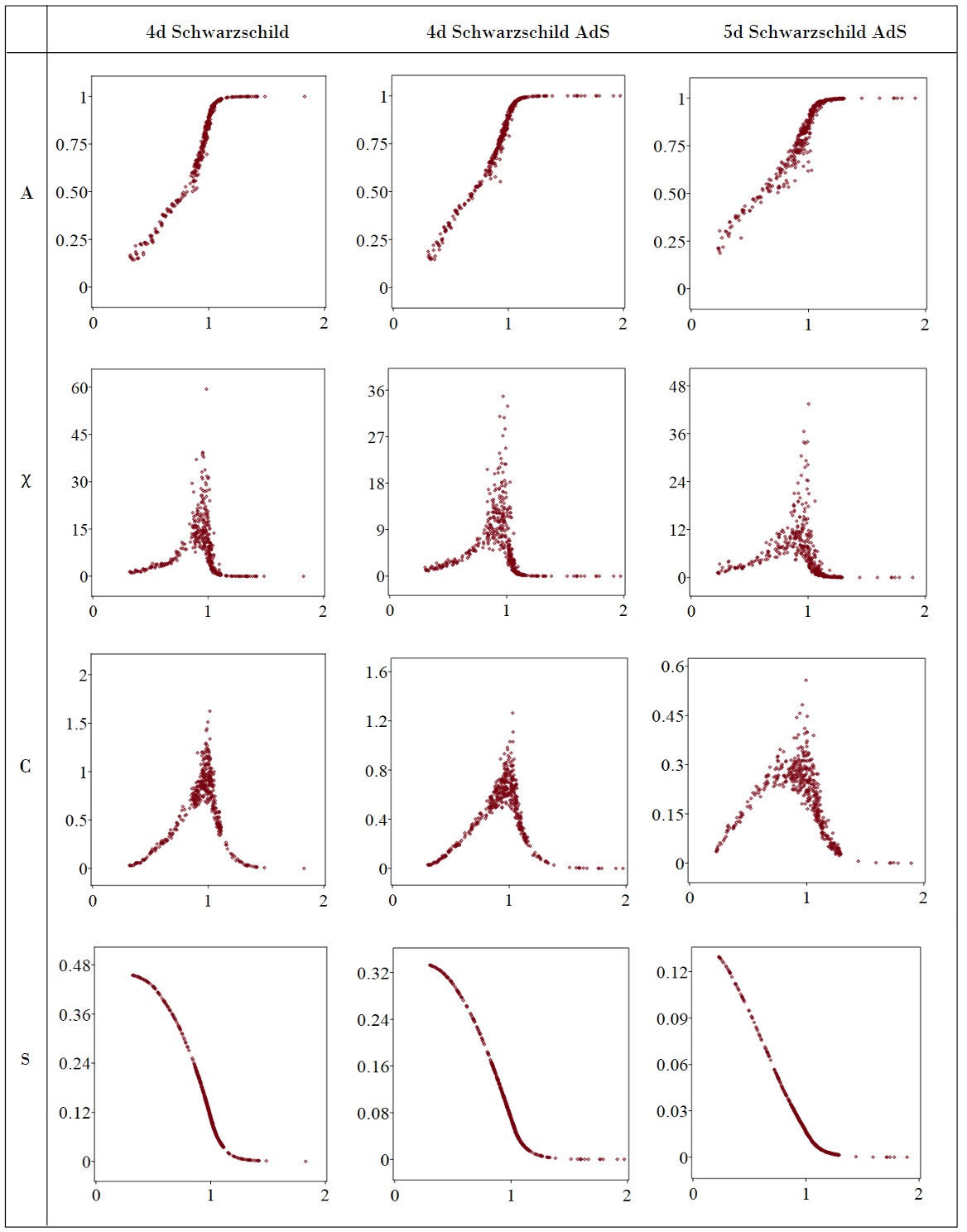}
\caption{\label{fig:spin-half-M}Plots of thermodynamic quantities for spin 1/2 models on Euclidean black holes  with respect to $\frac{M}{M_{c}}$. The critical masses for Schwarzschild, and 4d and 5d Schwarzschild AdS ($\Lambda = -4$) are 0.070, 0.072 and 0.023 respectively in Planck units. }
\end{center}
\end{figure*}

\begin{figure*}
\begin{center}
\includegraphics[width=\textwidth]{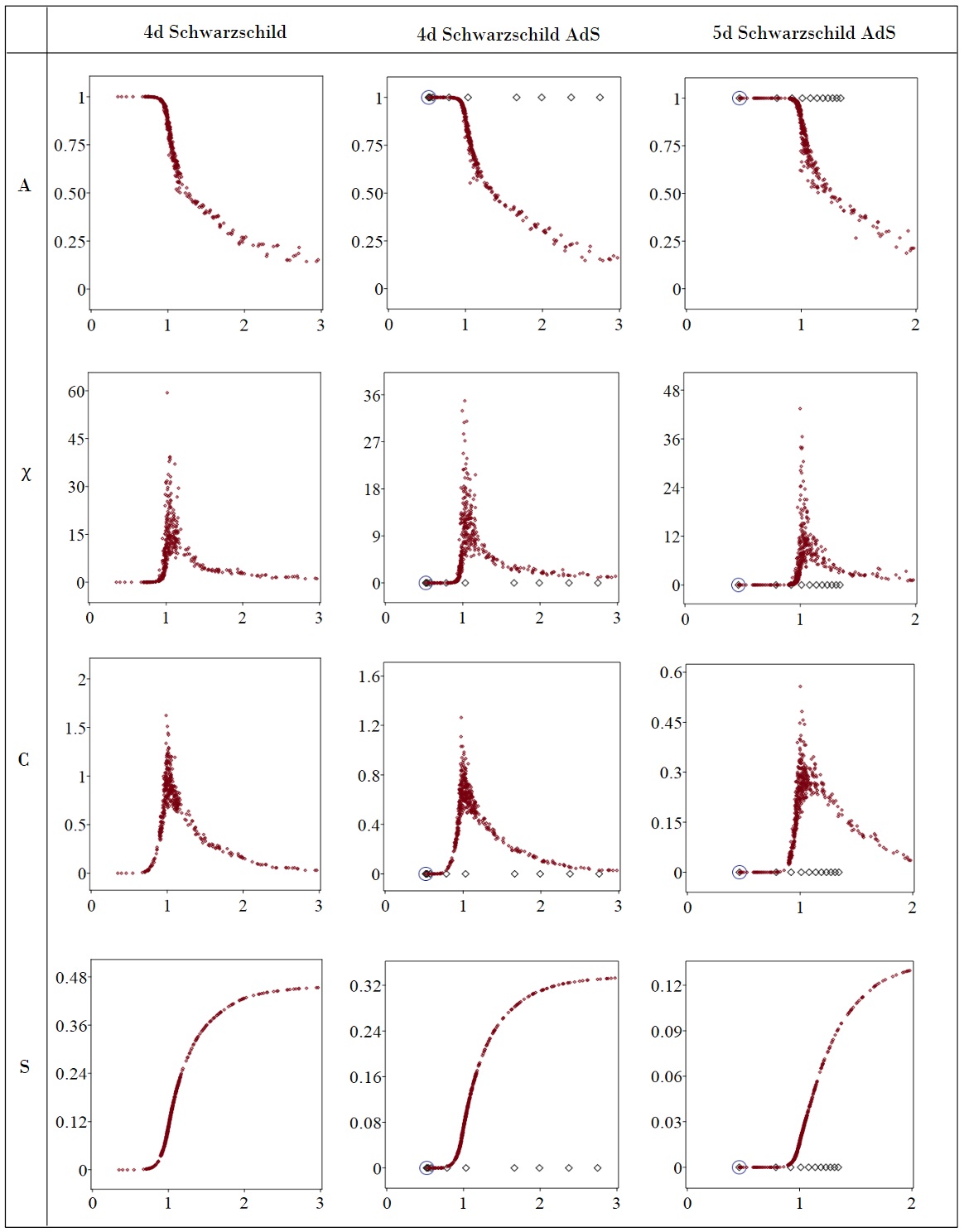}
\caption{\label{fig:spin-half-T}Plots of thermodynamic quantities for spin 1/2 models on Euclidean black holes with respect to $\frac{T}{T_{c}}$. The critical temperatures for Schwarzschild, 4d Schwarzschild AdS and 5d Schwarzschild AdS are 0.57, 0.61 and 0.80 respectively. For the AdS cases the point circled blue represents the black hole that corresponds to $T_{min}$, the black diamonds correspond to the large black holes, and the red dots correspond to the small black holes.}
\end{center}
\end{figure*}

\begin{figure*}
\begin{center}
\includegraphics[width=\textwidth]{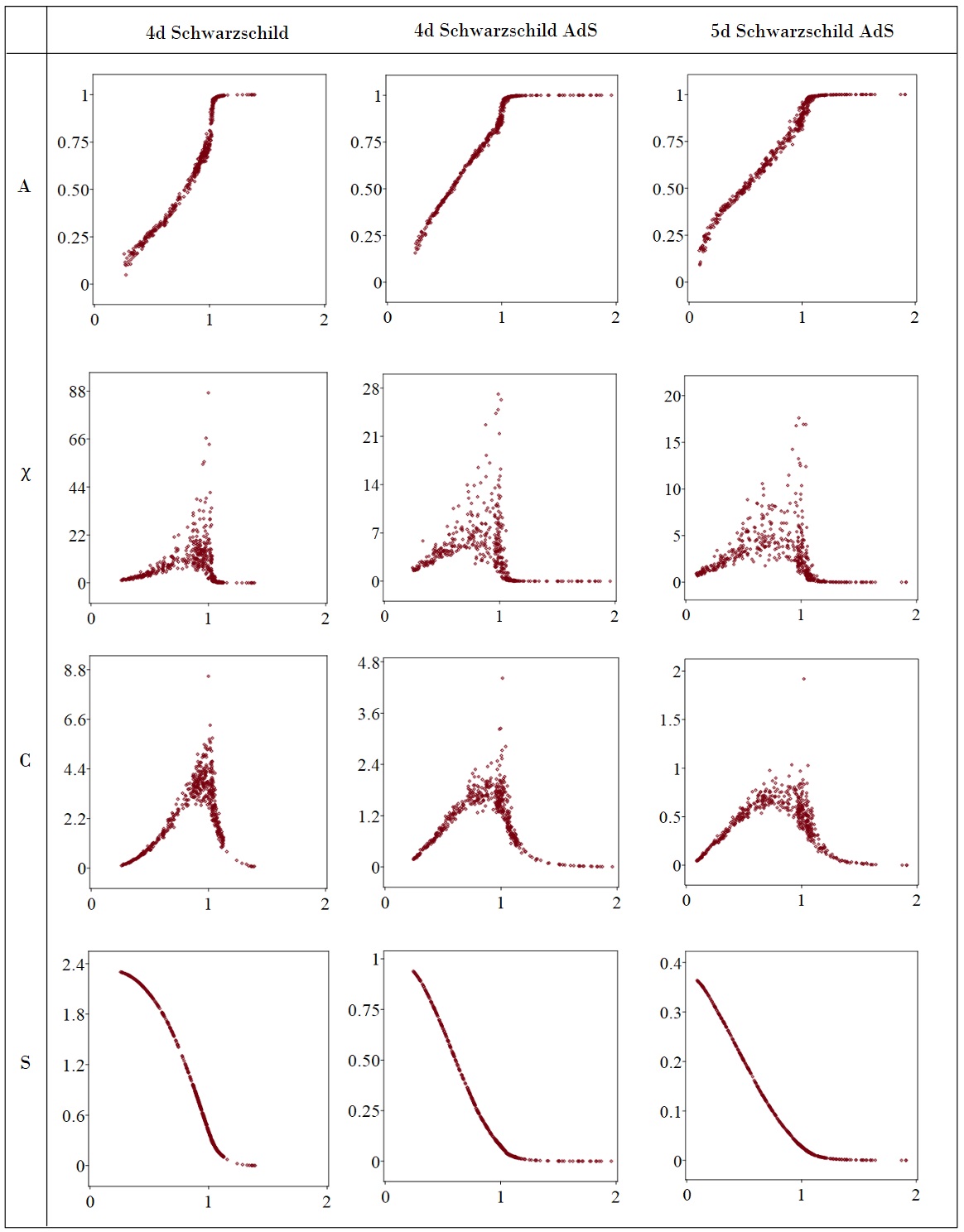}
\caption{\label{fig:spin-one-M}Plots of thermodynamic quantities for spin 1 models on Euclidean black holes with respect to $\frac{M}{M_{c}}$. The critical masses for Schwarzschild, 4d Schwarzschild AdS and 5d Schwarzschild AdS are 0.14, 0.17 and 0.065 respectively.}
\end{center}
\end{figure*}

\begin{figure*}
\begin{center}
\includegraphics[width=\textwidth]{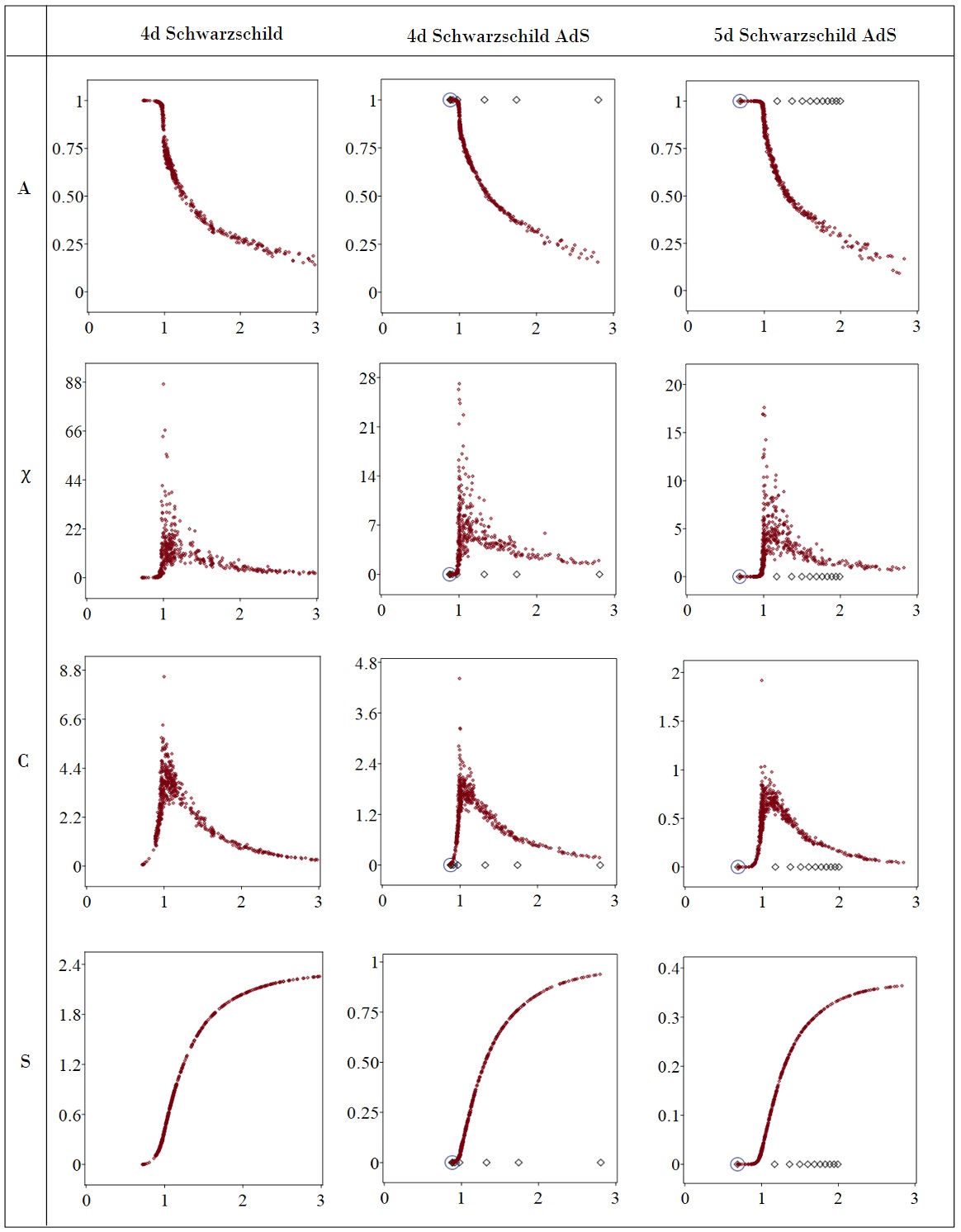}
\caption{\label{fig:spin-one-T}Plots of thermodynamic quantities for spin 1 models on Euclidean black holes with respect to $\frac{T}{T_{c}}$. The critical temperatures for Schwarzschild, 4d Schwarzschild AdS and 5d Schwarzschild AdS are 0.28, 0.36 and 0.53 respectively. For the AdS cases the point circled blue represents the black hole that corresponds to $T_{min}$, the black diamonds correspond to the large black holes, and the red dots correspond to the small black holes.}
\end{center}
\end{figure*}

Figures \ref{fig:spin-half-M} and \ref{fig:spin-half-T} exhibit thermodynamic quantities for the spin-1/2 case for the three Euclidean black hole backgrounds, as functions of $M/M_c$ and $T/T_c$ respectively; Figs. \ref{fig:spin-one-M} and \ref{fig:spin-one-T} show the same for spin-1. 

Evidence of a second order phase transition is apparent in the figures; specifically the specific heat  and susceptibility exhibit an approximate  divergence at the respective critical values of the masses, whereas the alignment indicates a transition from disorder to order as the mass increases. A curious feature is that the value of the critical masses is sub-Planckian (in the geometrized units $\hbar=c=G=1$).

The graphs of the thermodynamic variables as a function of $T/T_c$ contain two curves for the AdS cases; these are constructed by identifying the two possible black hole masses corresponding to a fixed $T$ value as in Fig. \ref{fig:AdS_TM}. To illustrate the procedure, consider for example the alignment graph as a function of $T/T_c$ for 4d Schwarzschild-AdS: a fixed $T$ identifies a large and a small black hole mass above the minimum value $T_{\rm min}$; we take the two alignment values for these two masses from the alignment vs. mass graph and plot these values at the fixed $T$; this process is repeated for all data. The critical temperature is obtained uniquely from the critical mass value. Thus for example, in the alignment vs. $T/T_c$ graphs in Fig. \ref{fig:spin-half-T}, the circled point corresponds to the minimum ($T_{\rm min}$) of the $T$ vs. $M$ curve in Fig.  \ref{fig:AdS_TM}; the black diamonds correspond to the large black holes, and the red dots correspond to the small black holes; in the range of $T$ shown, the spins are ordered (as functions of $T$) for the large black hole, but exhibit a transition to disorder for the small black hole. Similar results hold for all the other thermodynamic quantities: spins on the background of the small black hole exhibit a phase transition as functions of $T$, but spins on the background of the large black hole do not, at least for the range of $T$ values we were able to efficiently compute. Table 1 lists the critical masses and corresponding temperatures for all the cases studied.
\begin{table}
\begin{center}
\begin{tabular}{|p{3.7cm} p{0.35cm} p{0.35cm} p{0.35cm} p{0.35cm}|}
\hline
\multicolumn{1}{|c}{Background} & \multicolumn{2}{c}{Spin 1/2} & \multicolumn{2}{c|}{Spin 1}\\
\multicolumn{1}{|c}{} & $M_{c}$ & $T_{c}$ & $M_{c}$ & $T_{c}$ \\
\hline
4d Schwarzschild   & 0.070    &0.57 & 0.14 & 0.28\\
4d Schwarzschild AdS &   0.072  & 0.61 & 0.17 & 0.36   \\
5d Schwarzschild AdS & 0.023 & 0.80 & 0.065 & 0.53\\
\hline
\end{tabular}
\end{center}
\caption{Table of critical masses and critical temperatures for spin 1/2 and spin 1 models on the Euclidean black hole backgrounds considered.}
\end{table}

\begin{figure*}
\begin{center}
\includegraphics[width=\textwidth]{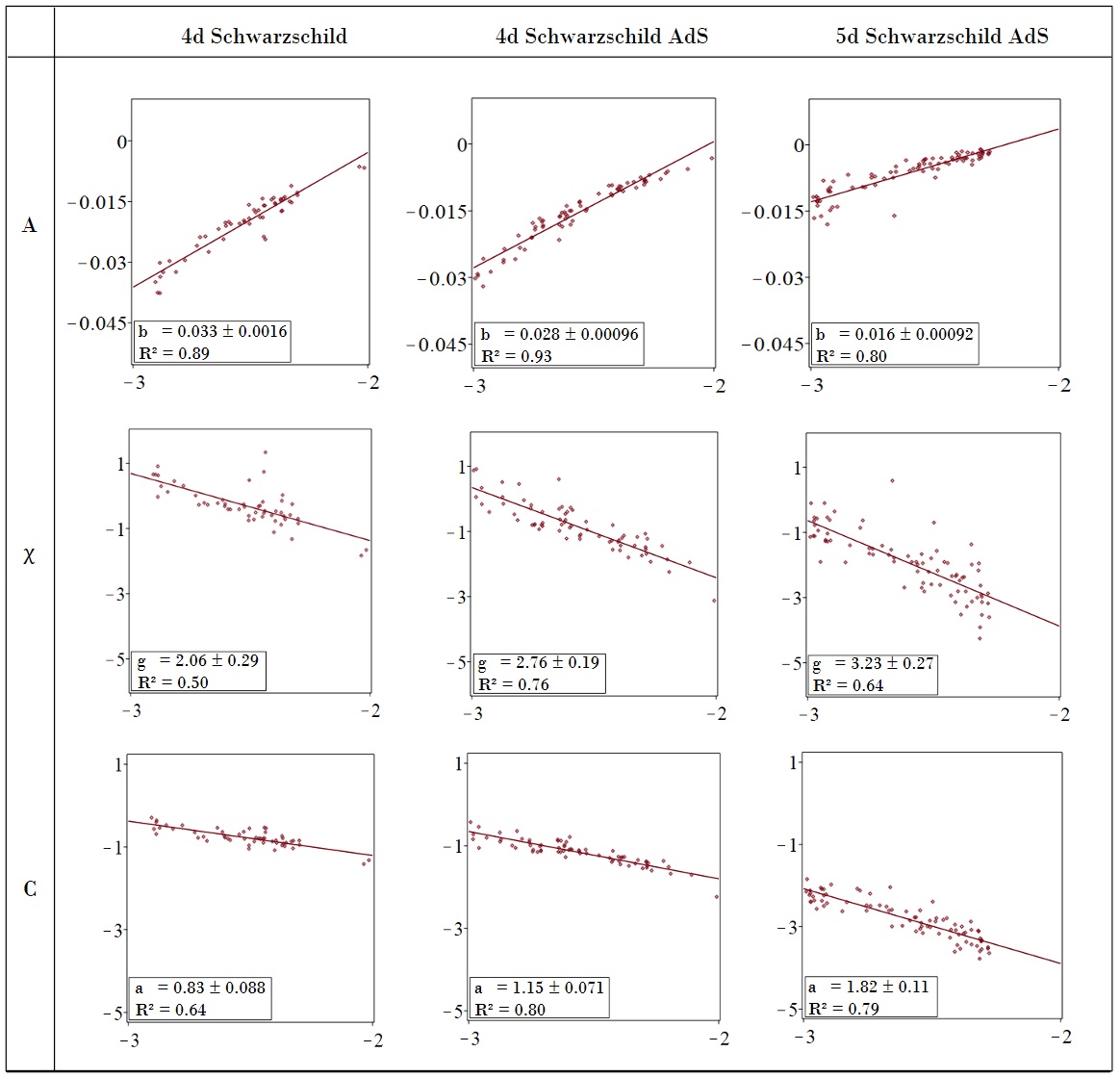}
\caption{\label{fig:CE-spin-half}Estimating critical exponents of thermodynamic quantities for spin 1/2 models on Euclidean black holes. The horizontal axis is ln($t$) and the vertical axis is the ln of the thermodynamic variable in the corresponding row.}
\end{center}
\end{figure*}

\begin{figure*}
\begin{center}
\includegraphics[width=\textwidth]{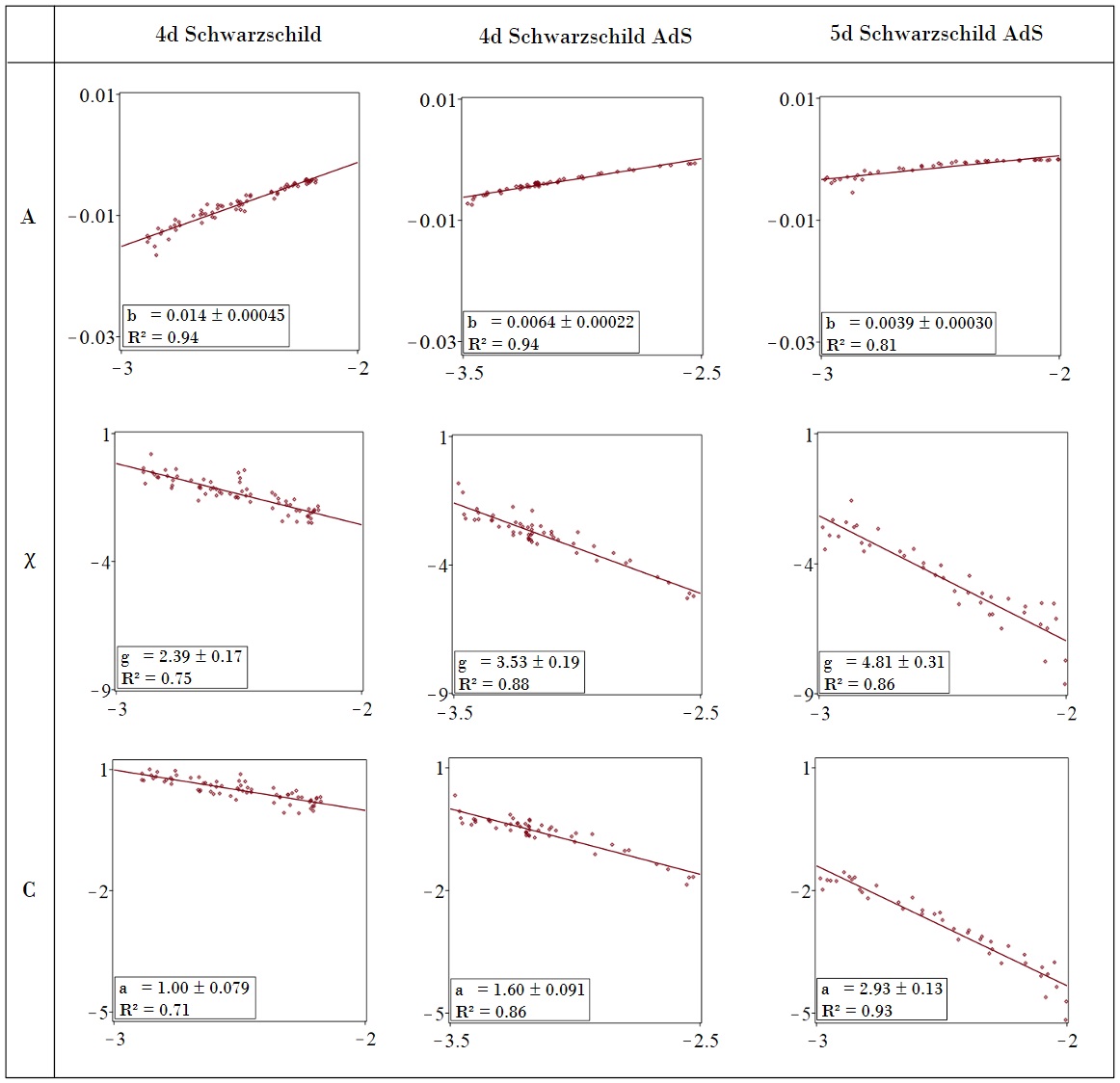}
\caption{\label{fig:CE-spin-one}Estimating critical exponents of thermodynamic quantities for spin 1 models on Euclidean black holes. The horizontal axis is ln($t$) and the vertical axis is the ln of the thermodynamic variable in the corresponding row.}
\end{center}
\end{figure*}

Lastly Figs. \ref{fig:CE-spin-half} and \ref{fig:CE-spin-one} 
are log-log graphs of the thermodynamics quantities for the spin-half and spin-one models as functions of the reduced temperature $t$, where  $t=1-\frac{T}{T_c}$; these figures exhibit linear behaviour and therefore power laws of the form  
\bea
\label{eq:ce00}
A &\approx& t^{b}, \nn\\  
\chi &\approx& t^{-g},  \nn\\ 
C &\approx & t^{-a},
\eea
together with least square fit estimates for the critical exponents $b,g$ and $a$. It is worth mentioning that for the Ising or Blume-Capel models, $\chi$ and $C$ from equation \eqref{eq:ce00} depend on $|t|$ and conventionally the critical exponents are calculated separately for $t>0$ and $t<0$ following which they are averaged; we use $t>0$ only since it corresponds to larger lattices and therefore more accurate results.
This completes the description of the results: our main observation is that spins on Euclidean black hole geometries in four and five dimensions undergo phase transitions as a function of the mass parameter associated to the geometries. Perhaps surprisingly, the phase transitions to order occur at approximately a tenth of Planck mass in 4d (Table 1), whereas an intuitive expectation might be a few Planck masses. Nevertheless a transition in this range is not surprising considering the Planck mass is the only natural scale in the spin model.  

\section{Discussion}

The model and results we have described may be viewed as falling in the broad area of quantum fields on curved spacetime \cite{birrell_davies, ParkerToms}. Most work in this area  is  concerned mainly with quantum fields on Lorentzian rather than Euclidean curved geometries; the exception is thermal QFT which is quantum fields on Wick rotated Minkowski space \cite{Sch,Nak,Sym1,Sym2}. The model and the calculations we have performed are new in that (to our knowledge) there are no prior studies of inhomogeneous spin models on Euclidean black holes. Our work may be viewed as exploring the extent to which a Euclidean black hole is equivalent to a thermal heat bath for matter degrees of freedom propagating on such a geometry; our work does not concern or provide comment on the use of such solutions as contributions to the path integral of Euclidean quantum gravity without matter, which is the arena for the Hawking-Page transition \cite{Hawking1}.  

The inhomogeneity of the spin models comes from the radial dependence of the 
coupling strength evident in the actions (\ref{eq:imsa4}) and (\ref{eq:imsb}). This variation of  coupling strength across the lattice means that the probability of a spin flip is dependent on radial position in the lattice; e.g. for Euclidean Schwarzschild the nearest-neighbour coupling in the $\tau$-direction varies as $M^3/[\rho_n (1-(\rho_n/4M)^2)]^4$, whereas the coupling in the  $\rho$-direction varies as  $\rho_n M$. This not only makes apparent the inhomogeneous nature of the model but it also confirms that for small/large $M$ the couplings are weak/strong which accounts for the observed disorder/order. The radial dependence of couplings also means that at the transition point from order to disorder does not have the scale invariance of homogeneous models observed across the lattice. This in turn indicates that the assignment of the term ``second-order transition" comes with qualifications despite the linearity observed in the log-log graphs in Figs. \ref{fig:CE-spin-half} and \ref{fig:CE-spin-one}. 

Numerical accuracy of our results can be improved. This may be accomplished by an MC algorithm that is more efficient than the single spin flip one used in this work. However it is worth  noting that most MC algorithms - such as Wolff's algorithm \cite{LandauBinder, Pang} and that proposed in \cite{LandauWang} - are designed to study spins on a flat Euclidean background for which temperature is an external parameter and is not related to lattice size. Therefore the MC algorithm being used must first be redesigned to apply to the type of inhomogeneous models that arise on black hole backgrounds. Another way to improve accuracy is via the finite size scaling technique \cite{Landau,LandauBinder,Thijssen}, where successively finer lattices are used to approach the continuum limit. 

Beyond improving MC methods, there are several directions for extending this work on spin models  on Euclidean black hole backgrounds: exploring thermodynamic properties while varying both mass and cosmological constant; Reissner-N\"{o}rdstrom black holes and other black holes, including  extremal cases; higher spin; continuum of scalar field values within a given range rather than discrete; and adding mass and other self-interaction terms.

{\bf Acknowledgements} This work was supported by the Natural Science and Engineering Research Council of Canada.

\bibliography{spinBH}

\end{document}